\documentclass[12pt,a4paper]{article}
	
\usepackage{a4wide}
\usepackage{amsmath}
\usepackage{amssymb}
\usepackage{amsfonts}
\usepackage{epsfig}
\usepackage{exscale}
\usepackage{float}
\usepackage{bbm}
\usepackage[numbers,sort&compress]{natbib}

\newcommand{\Z}{{\mathbb{Z}}}
\newcommand{\R}{{\mathbb{R}}}

\newcommand{\p}{\partial}

\makeatletter
\@addtoreset{equation}{section}
\makeatother

\title{Self-adjoint Extensions for Confined Electrons: from a Particle
in a Spherical Cavity to the Hydrogen Atom in a Sphere and on a Cone}

\author{M.\ H.\ Al-Hashimi and U.-J.\ Wiese \\ \\
Albert Einstein Center for Fundamental Physics \\
Institute for Theoretical Physics, Bern University \\
Sidlerstrasse 5, CH-3012 Bern, Switzerland \\ \\}

\begin{document} 

\maketitle

\vspace{-1cm}

\begin{abstract} \normalsize

In a recent study of the self-adjoint extensions of the Hamiltonian of a
particle confined to a finite region of space, in which we generalized the 
Heisenberg uncertainty relation to a finite volume, we encountered bound states 
localized at the wall of the cavity. In this paper, we study this situation in
detail both for a free particle and for a hydrogen atom centered in a 
spherical cavity. For appropriate values of the self-adjoint extension 
parameter, the bound states localized at the wall resonate with the standard 
hydrogen bound states. We also examine the accidental symmetry generated by the 
Runge-Lenz vector, which is explicitly broken in a spherical cavity with general
Robin boundary conditions. However, for specific radii of the confining sphere, 
a remnant of the accidental symmetry persists. The same is true for an electron 
moving on the surface of a finite circular cone, bound to its tip by a $1/r$ 
potential.

\end{abstract}

\newpage
 
\section{Introduction}

The confinement of atoms or molecules in a finite region of space is an 
important subject, for example, in the context of quantum dots, encapsulation in
fullerenes, or other aspects of nanotechnology \cite{Sab09}. Confinement in a 
cavity has also been used to mimic the effects of high pressure, e.g., in the 
modeling of white dwarf stars. A hydrogen atom at the center of a spherical 
cavity was first studied by Michels, de Boer, and Bijl in 1937 \cite{Mic37} in 
order to model hydrogen at high pressure, as well as by Sommerfeld and Welker
in 1938 \cite{Som38}, and also in an extended body of subsequent work 
\cite{Gro46,Wig54,Fow84,Fro87}. Most of these works use the standard Dirichlet 
boundary condition with a vanishing wave function at the cavity wall, while 
some use Neumann boundary conditions, i.e.\ a vanishing gradient of the wave 
function perpendicular to the wall. A notable exception is \cite{Sch00} where 
the most general so-called Robin boundary condition has been considered.
The finite-volume remnants of the accidental symmetry associated with the 
Runge-Lenz vector have been studied in \cite{Pup98,Sch00,Pup02}. In 
this paper, we investigate a free particle as well as a hydrogen atom in
a spherical cavity with Robin boundary conditions in detail. In particular, we
investigate the spectrum as a function of the self-adjoint extension parameter
that characterizes the boundary condition. We also extend our earlier study of
accidental symmetries of an electron confined to the surface of a cone and bound
to its tip by a $1/r$ potential \cite{AlH08} to a cone of finite extent. Just as
for the hydrogen atom in a spherical cavity \cite{Sch00}, we find a remnant of 
the accidental degeneracy on a finite cone for particular values of the radius 
and of the self-adjoint extension parameter. Atoms encapsulated in fullerenes 
may resonate with the confining shell \cite{Pus93,Wen93}. This has been modeled 
with attractive potentials near the boundary \cite{Con99,Con00}. As we will see,
describing the physical properties of the perfectly reflecting cavity wall by a 
self-adjoint extension parameter also gives rise to cavity resonances.

In the standard quantum mechanics textbooks, the differences between
Hermiticity and self-adjointness are rarely emphasized. As a consequence, even
the simple system of a particle in a box with perfectly reflecting walls is 
usually not discussed in its most general form. Indeed, there is a self-adjoint 
extension parameter $\gamma(\vec x)$ that specifies the physical properties of 
the reflecting wall \cite{Bon01}. The corresponding Robin boundary condition 
takes the form
\begin{equation}
\label{bcdot}
\gamma(\vec x) \Psi(\vec x) + \vec n(\vec x) \cdot \vec \nabla \Psi(\vec x) = 0,
\quad \vec x \in \p \Omega,
\end{equation}
where $\p \Omega$ is the boundary of a spatial region $\Omega$ and 
$\vec n(\vec x)$ is the unit-vector normal to the surface. In the usual textbook
discussions one sets $\gamma(\vec x) = \infty$, which implies that the wave 
function $\Psi(\vec x)$ vanishes at the boundary. However, this is not necessary
because eq.(\ref{bcdot}) always guarantees that
\begin{equation}
\vec n(\vec x) \cdot \vec j(\vec x) = 0, \quad \vec x \in \p \Omega,
\end{equation}
i.e.\ it ensures that the component of the probability current density 
\begin{equation}
\vec j(\vec x,t) = \frac{1}{2 M i}
\left[\Psi(\vec x,t)^* \vec \nabla \Psi(\vec x,t) -
\vec \nabla \Psi(\vec x,t)^* \Psi(\vec x,t)\right],
\end{equation}
normal to the surface vanishes. Hence, together with the continuity equation
\begin{equation}
\p_t \rho(\vec x,t) + \vec \nabla \cdot \vec j(\vec x,t) = 0, \quad
\rho(\vec x,t) = |\Psi(\vec x,t)|^2,
\end{equation}
the boundary condition eq.(\ref{bcdot}) ensures probability conservation.
This is the key feature that guarantees the self-adjointness (rather than just
the Hermiticity) of the Hamiltonian
\begin{equation}
H = \frac{{\vec p \,}^2}{2 M} + V(\vec x) = - \frac{1}{2 M} \Delta + V(\vec x).
\end{equation}
In order to discuss the issue of Hermiticity, let us consider
\begin{eqnarray}
\langle\chi|H|\Psi\rangle&=&
\int_\Omega d^3x \ \chi(\vec x)^* 
\left[- \frac{1}{2 M} \Delta + V(\vec x)\right] \Psi(\vec x) \nonumber \\
&=&\int_\Omega d^3x \ 
\left[\frac{1}{2 M} \vec \nabla \chi(\vec x)^* \cdot \vec \nabla \Psi(\vec x) +
\chi(\vec x)^* V(\vec x) \Psi(\vec x) \right] \nonumber \\
&-&\frac{1}{2 M}
\int_{\p \Omega} d\vec n \cdot \chi(\vec x)^* \vec \nabla \Psi(\vec x)
\nonumber \\
&=&\int_\Omega d^3x \ \left\{\left[- \frac{1}{2 M} \Delta + V(\vec x)\right]
\chi(\vec x)^* \right\} \Psi(\vec x) \nonumber \\
&+&\frac{1}{2 M} \int_{\p \Omega} d\vec n \cdot
\left[\vec \nabla \chi(\vec x)^* \Psi(\vec x) - 
\chi(\vec x)^* \vec \nabla \Psi(\vec x)\right] \nonumber \\
&=&\langle\Psi|H|\chi\rangle^* + \frac{1}{2 M} \int_{\p \Omega} d\vec n \cdot
\left[\vec \nabla \chi(\vec x)^* \Psi(\vec x) - 
\chi(\vec x)^* \vec \nabla \Psi(\vec x)\right].
\end{eqnarray}
The Hamiltonian is Hermitean if
\begin{equation}
\label{symmetricHdot}
\int_{\p \Omega} d\vec n \cdot
\left[\vec \nabla \chi(\vec x)^* \Psi(\vec x) - 
\chi(\vec x)^* \vec \nabla \Psi(\vec x)\right] = 0.
\end{equation}
Using the boundary condition eq.(\ref{bcdot}), the integral in 
eq.(\ref{symmetricHdot}) reduces to
\begin{equation}
\int_{\p \Omega} d^2x \left[\vec n(\vec x) \cdot \vec \nabla \chi(\vec x)^* + 
\gamma(\vec x) \chi(\vec x)^*\right] \Psi(\vec x) = 0.
\end{equation}
Since $\Psi(\vec x)$ itself can take arbitrary values at the boundary, the
Hermiticity of $H$ requires that
\begin{equation}
\label{bcdual}
\vec n(\vec x) \cdot \vec \nabla \chi(\vec x) + \gamma(\vec x)^* \chi(\vec x) = 
0.
\end{equation}
Self-adjointness (rather than just Hermiticity) of the Hamiltonian requires that
the domains $D(H)$ and $D(H^\dagger)$ agree with each other \cite{Neu32,Ree75}. 
The domain $D(H)$
consists of those twice-differentiable and square-integrable wave functions that
obey the boundary condition (\ref{bcdot}), while the domain $D(H^\dagger)$ 
consists of the functions obeying eq.(\ref{bcdual}). For 
$\gamma(\vec x) \in \R$, the two conditions agree and thus, 
$D(H^\dagger) = D(H)$. Consequently, there is a 1-parameter family (parameterized
by $\gamma(\vec x)$) that specifies the self-adjoint extensions of the 
Hamiltonian in the finite domain $\Omega$. In this paper, we consider a 
spherical cavity $\Omega$ with a constant (and thus rotation invariant) 
self-adjoint extension parameter $\gamma$ that characterizes the physical
property of the perfectly reflecting boundary.

The paper is organized as follows. In section 2 we investigate a particle 
confined to a spherical cavity with general reflecting boundary 
conditions. Besides determining the energy spectrum, we discuss the 
corresponding Heisenberg uncertainty relation generalized to a finite volume. 
In section 3, the discussion is extended to the hydrogen problem, also 
considering the fate of the accidental symmetry generated by the Runge-Lenz 
vector, which is no longer a Hermitean operator. We also consider
a resonance phenomenon that arises for particular sizes of the cavity for 
negative values of the self-adjoint extension parameter $\gamma$. In section 4
we extend the discussion to an electron confined to the surface of a finite 
circular cone, and bound to its tip by a $1/r$ potential. Finally, section 5 
contains our conclusions.

\section{Particle in a Spherical Cavity with General Reflecting Boundaries}

As a preparation for the hydrogen problem, in this section we consider a 
``free'' particle in a spherical cavity with general reflecting boundary 
conditions specified by the self-adjoint extension parameter $\gamma \in \R$.

\subsection{Energy spectrum}

Let us consider the Hamiltonian of a free particle of mass $M$,
\begin{equation}
H = - \frac{1}{2 M} \Delta = - \frac{1}{2 M} \left(\p_r^2 + \frac{2}{r} \p_r
- \frac{\vec L \, ^2}{r^2}\right),
\end{equation}
with angular momentum $\vec L$ in a spherical cavity of radius $R$. As usual,
the wave function can be factorized as
\begin{equation}
\Psi(\vec x) = \psi_{k l}(r) Y_{lm}(\theta,\varphi),
\end{equation}
where the angular dependence is described by the spherical harmonics 
$Y_{lm}(\theta,\varphi)$. The radial wave function obeys
\begin{equation}
- \frac{1}{2 M} \left(\p_r^2 + \frac{2}{r} \p_r
- \frac{l(l + 1)}{r^2}\right) \psi_{k l}(r) = E \psi_{k l}(r), \quad
E = \frac{k^2}{2 M}.
\end{equation}
For positive energy the normalizable wave function is given by
\begin{equation}
\psi_{k l}(r) = A j_l(k r).
\end{equation}
For a spherical cavity, the most general perfectly reflecting boundary 
condition of eq.(\ref{bcdot}) takes the form
\begin{equation}
\gamma \psi_{k l}(R) + \p_r  \psi_{k l}(R) = 0,
\end{equation}
The energy spectrum is thus determined from
\begin{equation}
\gamma j_l(k R) + \p_r  j_l(k R) = 
\left(\gamma + \frac{l}{R} \right) j_l(k R) - k j_{l+1}(k R) = 0,
\end{equation}
which is a transcendental equation for $k \in \R$. 

\begin{figure}[tbh]
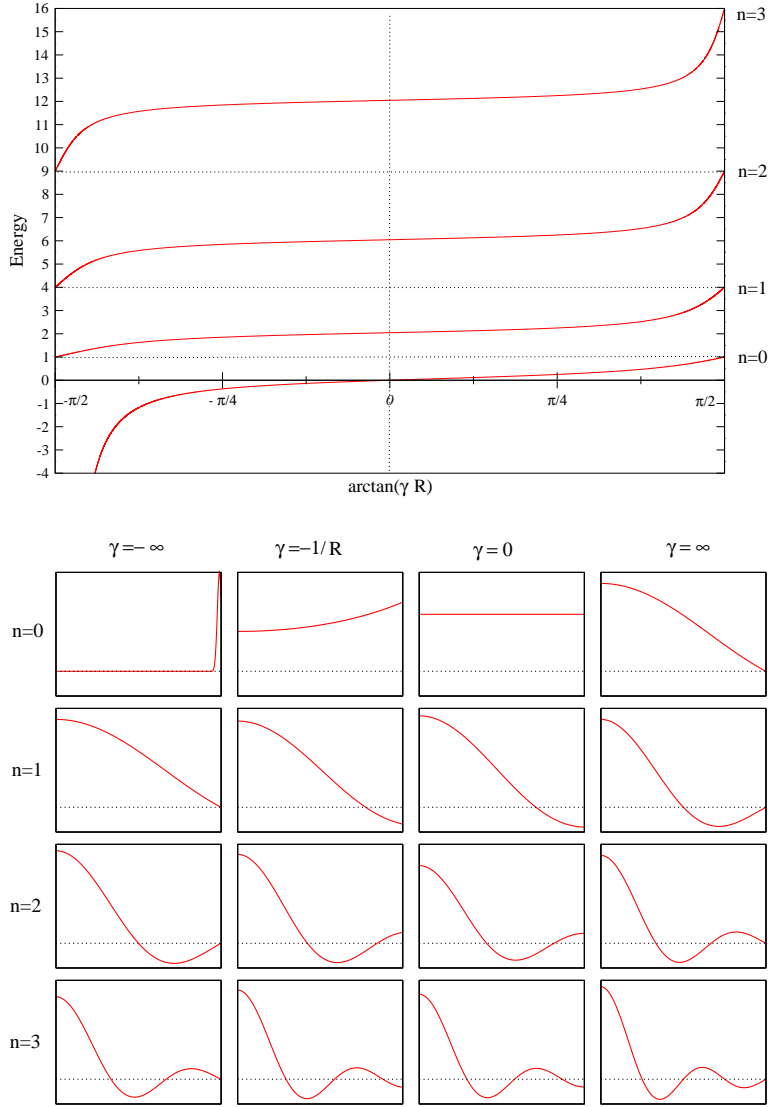

\begin{center}
\epsfig{file=hydfig1.eps,width=10cm} \vskip0.5cm
\epsfig{file=hydfig2.eps,width=10cm}
\end{center}
\caption{\it Top: Spectrum of $l = 0$ states for a particle in a spherical 
cavity with general Robin boundary conditions as a function of the self-adjoint 
extension parameter $\gamma$, rescaled to $\arctan(\gamma R)$. The energy 
is measured in units of $\pi^2/2 M R^2$. The dotted lines represent the spectrum
for $\gamma = \infty$. Bottom: Wave functions of the four lowest $l = 0$ states 
with $n = 0, 1, 2$ and $3$ for $\gamma = \infty,0,-1/R$, and $-\infty$.}
\label{spectrum0}
\end{figure}

\begin{figure}[tbh]
\begin{center}
\epsfig{file=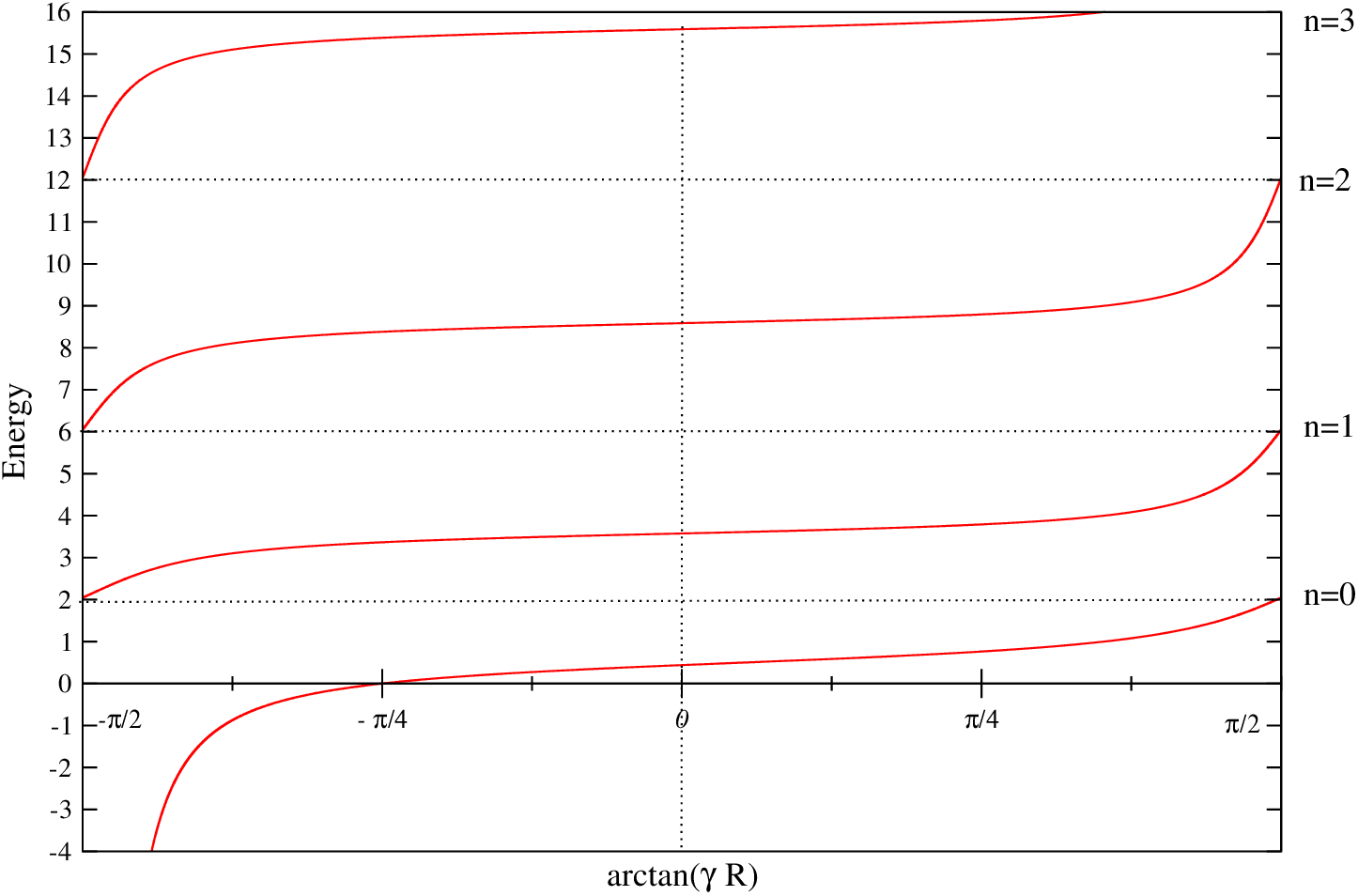,width=10cm} \vskip0.5cm
\epsfig{file=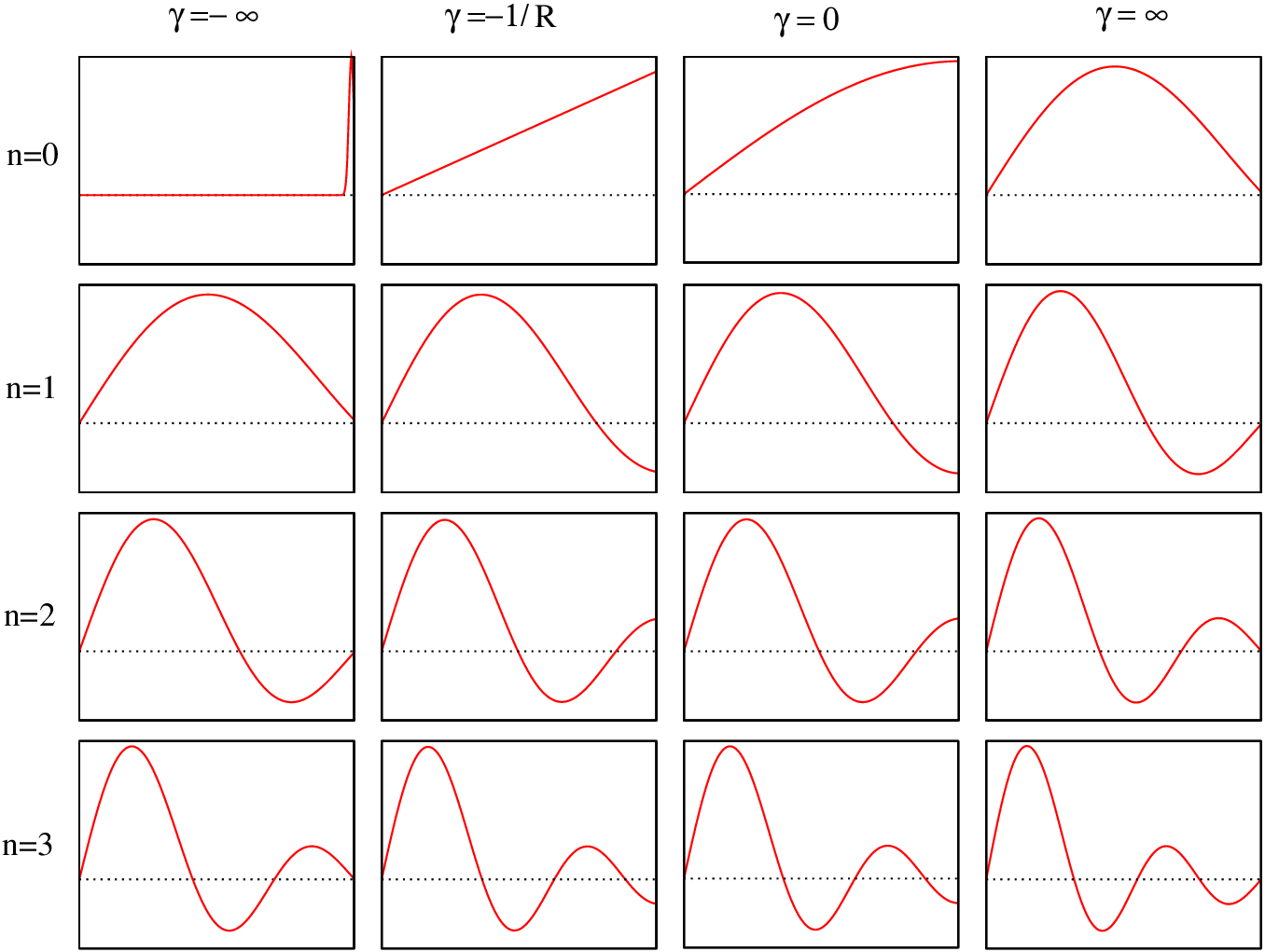,width=10cm}
\end{center}
\caption{\it Top: Spectrum of $l = 1$ states for a particle in a spherical 
cavity with general Robin boundary conditions as a function of the self-adjoint 
extension parameter $\gamma$, rescaled to $\arctan(\gamma R)$. The energy 
is measured in units of $\pi^2/2 M R^2$. The dotted lines represent the spectrum
for $\gamma = \infty$. Bottom: Wave functions of the four lowest $l = 1$ states 
with $n = 0, 1, 2$ and 3 for $\gamma = \infty,0,-1/R$, and $-\infty$.}
\label{spectrum1}
\end{figure}

First, let us consider s-states with $l = 0$. For $\gamma \rightarrow \infty$ 
the boundary condition reduces to the textbook case $\psi(R) = 0$. The 
corresponding energies are then given by
\begin{equation}
E_{n0}(\gamma \rightarrow \infty) = \frac{(n + 1)^2 \pi^2}{2 M R^2}, \quad 
n = 0, 1, 2, \dots
\end{equation}
Similarly, for $\gamma = 1/R$ one obtains 
\begin{equation}
E_{n0}(\gamma = 1/R) = \frac{(n + \frac{1}{2})^2 \pi^2}{2 M R^2}, \quad 
n = 0, 1, 2, \dots
\end{equation}
For arbitrary $l \geq 1$ one can show that
\begin{equation}
E_{nl}(\gamma = (l+1)/R) = E_{n,l-1}(\gamma \rightarrow \infty), \quad 
n = 0, 1, 2, \dots,
\end{equation}
in particular,
\begin{equation}
E_{n1}(\gamma = 2/R) = \frac{(n + 1)^2 \pi^2}{2 M R^2}, \quad 
n = 0, 1, 2, \dots
\end{equation}
At $\gamma = 0$ (and for general $l$ at $\gamma = - l/R$), the ground state has
zero energy with the radial wave function given by
\begin{equation}
\psi(r) = \sqrt{\frac{2 l + 3}{R^3}} \left(\frac{r}{R}\right)^l,
\end{equation}
while the excited states have energies given by
\begin{equation}
E_{n+1,l}(\gamma = - l/R) = E_{n,l+1}(\gamma \rightarrow \infty), \quad 
n = 0, 1, 2, \dots
\end{equation}
Interestingly, for $\gamma < - l/R$, 
there are negative energy states, although the particle seems to have only 
kinetic energy. This is a consequence of the general Robin boundary conditions. 
While they are perfectly reflecting for positive energy states, they may still 
bind negative energy states to the wall. The negative energy states simply 
follow by analytic continuation of $k$ to $i k$. For 
$\gamma \rightarrow - \infty$ there is a bound state for each angular momentum 
$l$, with the energy
\begin{equation}
E_{0l}(\gamma \rightarrow - \infty) \rightarrow - \frac{\gamma^2}{2 M}.
\end{equation}
The other states remain at positive energies and, in fact, agree with those
at $\gamma \rightarrow \infty$. In particular, for $l = 0$, one then obtains
\begin{equation}
E_{n0}(\gamma \rightarrow - \infty) = \frac{n^2 \pi^2}{2 M R^2}, \quad 
n = 1, 2, 3, \dots
\end{equation}
The energy spectrum for $l = 0$ and the corresponding wave functions of the
states with $n = 0, 1, 2, 3$ are illustrated in figure \ref{spectrum0}. 
Analogous results are shown in figure \ref{spectrum1} for $l = 1$.

\subsection{Generalized Uncertainty Relation}

As we have seen, due to the general Robin boundary conditions, a particle
confined to a spherical cavity, which is otherwise free and thus seems to have 
only kinetic energy, may still have negative energy eigenvalues. This apparently
contradicts the Heisenberg uncertainty relation, because the necessarily finite
uncertainty of the position of a confined particle seems to imply a positive
kinetic energy. However, this is not necessarily the case because the standard 
Heisenberg
uncertainty relation was derived for an infinite volume and thus does not apply
in a finite cavity. Recently, we have derived a generalized uncertainty relation
valid in an arbitrarily shaped finite region $\Omega$ with the unit-vector 
$\vec n$ perpendicular to the boundary $\p \Omega$ \cite{AlH12}
\begin{equation}
\label{uncertainty}
2 M E_n = \langle {\vec p \,}^2 \rangle \geq \left(\frac{3 + 
\langle \vec n \rangle \cdot \langle \vec x \rangle - 
\langle \vec n \cdot \vec x \rangle}{2 \Delta x}\right)^2 + 
\langle \gamma \rangle +
\frac{\langle \vec n \rangle^2}{4},
\end{equation}
where we have defined
\begin{eqnarray}
&&\langle \vec n \cdot \vec x \rangle =
\int_{\p \Omega} d\vec n \cdot \vec x \rho(\vec x), \nonumber \\
&&\langle \gamma \rangle = 
\int_{\p \Omega} d^2x \ \gamma(\vec x) \rho(\vec x), \nonumber \\
&&\langle \vec n \rangle =
\int_{\p \Omega} d\vec n \ \rho(\vec x), \quad 
\rho(\vec x) = |\Psi(\vec x)|^2.
\end{eqnarray}
In the infinite volume limit, for localized states (with momentum expectation
value $\langle \vec p \rangle = 0$), the probability density 
vanishes at infinity and one obtains $\langle \vec n \cdot \vec x \rangle = 0$,
$\langle \gamma \rangle = 0$, and $\langle \vec n \rangle = 0$, such that one
recovers the usual Heisenberg uncertainty relation in three dimensions
\begin{equation}
\langle {\vec p \,}^2 \rangle \geq \left(\frac{3}{2 \Delta x}\right)^2 \
\Rightarrow \ \Delta x \Delta p \geq \frac{3}{2}.
\end{equation}

Let us now consider the generalized uncertainty relation in the context of the
particle confined to a spherical cavity. In this case, for the energy
eigenstates we obtain
\begin{equation}
\langle \vec n \cdot \vec x \rangle = R^3 |\psi_{k l}|^2, \quad
\langle \gamma \rangle = \gamma R^2 |\psi_{k l}|^2, \quad
\langle \vec n \rangle = 0.
\end{equation}
We are particularly interested in the zero-energy states, i.e.\ 
$k \rightarrow 0$, which arise for $\gamma = - l/R$. In that case, one obtains
\begin{equation}
\langle \vec n \cdot \vec x \rangle = 2 l + 3, \quad
\langle \gamma \rangle = (2 l + 3) \frac{\gamma}{R} = - \frac{l(2 l + 3)}{R^2}, 
\quad \Delta x = \sqrt{\frac{2 l + 3}{2 l + 5}} R.
\end{equation}
Inserting this in the generalized uncertainty relation eq.(\ref{uncertainty}),
yields
\begin{equation}
0 \geq \frac{1}{R^2}\left(\frac{l^2(2 l + 5)}{2 l + 3} - l(2 l + 3)\right).
\end{equation}
Indeed, this inequality is satisfied for all values of $l>0$, because
\begin{equation}
0 > l(2 l + 5) - (2 l + 3)^2 = - 2 l^2 - 7 l - 9.
\end{equation}
For $l = 0$ the inequality is saturated and hence, the corresponding wave 
function, which is constant for $\gamma = 0$, represents a minimal uncertainty
wave packet in the finite volume.

\section{Hydrogen Atom in a Spherical Cavity with General Reflecting Boundaries}

In this section we consider an electron bound to a proton that is localized at
the center of a spherical cavity with general reflecting boundary conditions,
again specified by the self-adjoint extension parameter $\gamma \in \R$.
\begin{figure}[tbh]
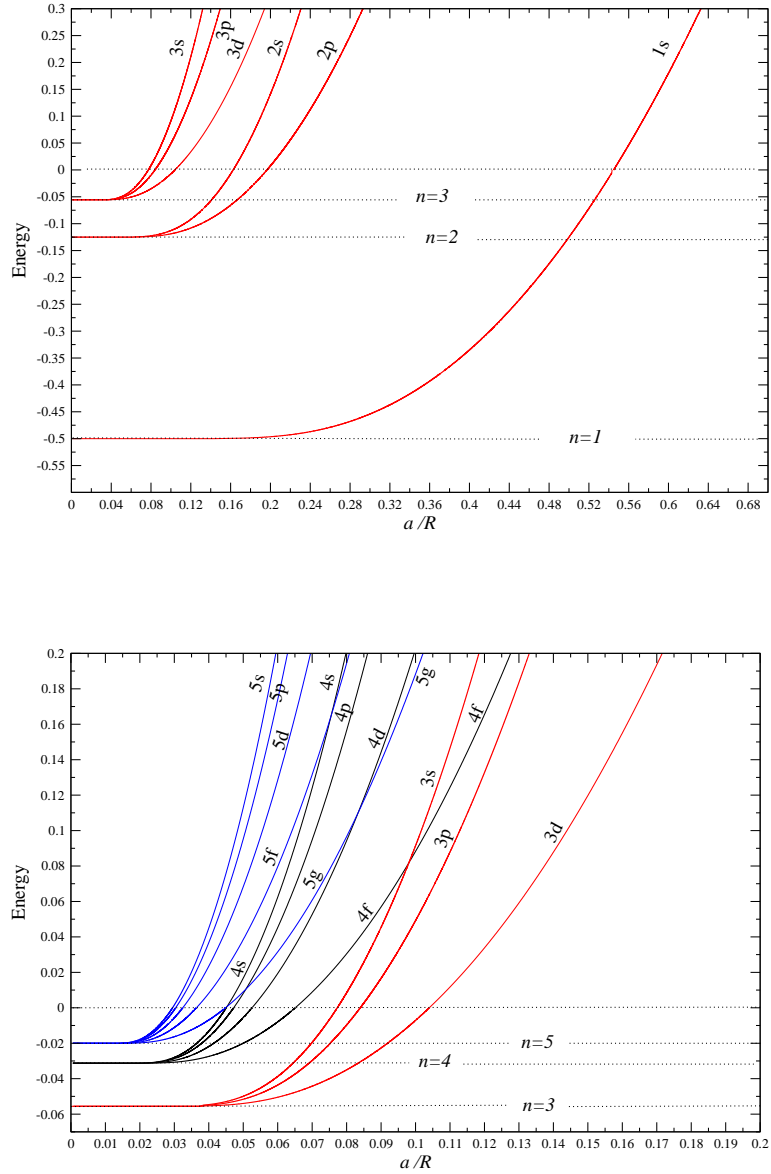

\begin{center}
\epsfig{file=hydfigA.eps,width=10cm} \vskip1.5cm
\epsfig{file=hydfigB.eps,width=10cm}
\end{center}
\caption{\it Spectrum of a hydrogen atom centered in a spherical cavity with 
the standard Dirichlet boundary condition (i.e.\ $\gamma = \infty$) as a 
function of $a/R$. The dotted lines represent the spectrum of the infinite 
system. Top: States with quantum numbers $n = 1, 2, 3$. Bottom: States with 
quantum numbers $n = 3, 4, 5$. The energy is given in units of $M e^4$.} 
\label{specinf}
\end{figure}
\begin{figure}[tbh]
\begin{center}
\epsfig{file=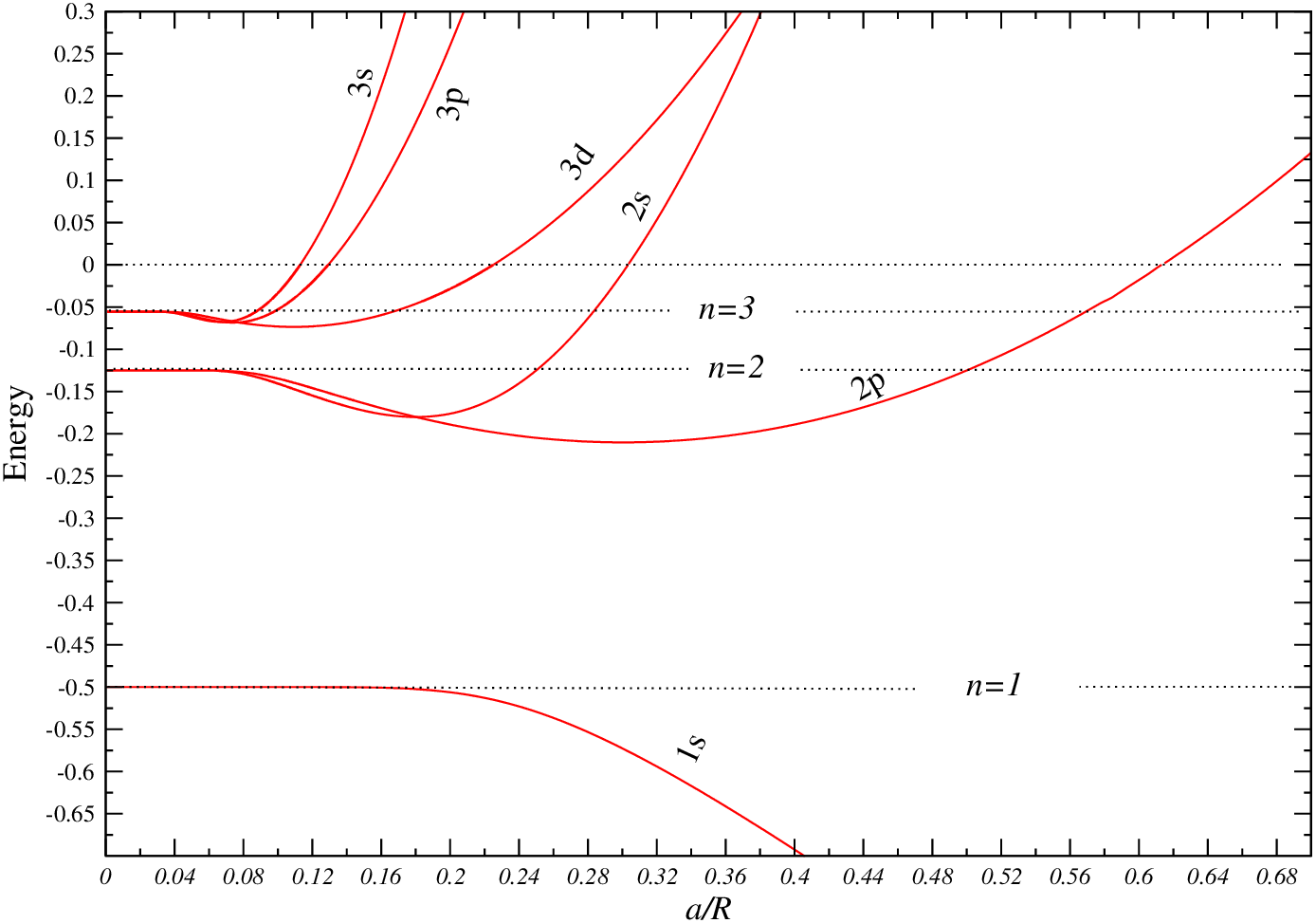,width=10cm} \vskip1.5cm
\epsfig{file=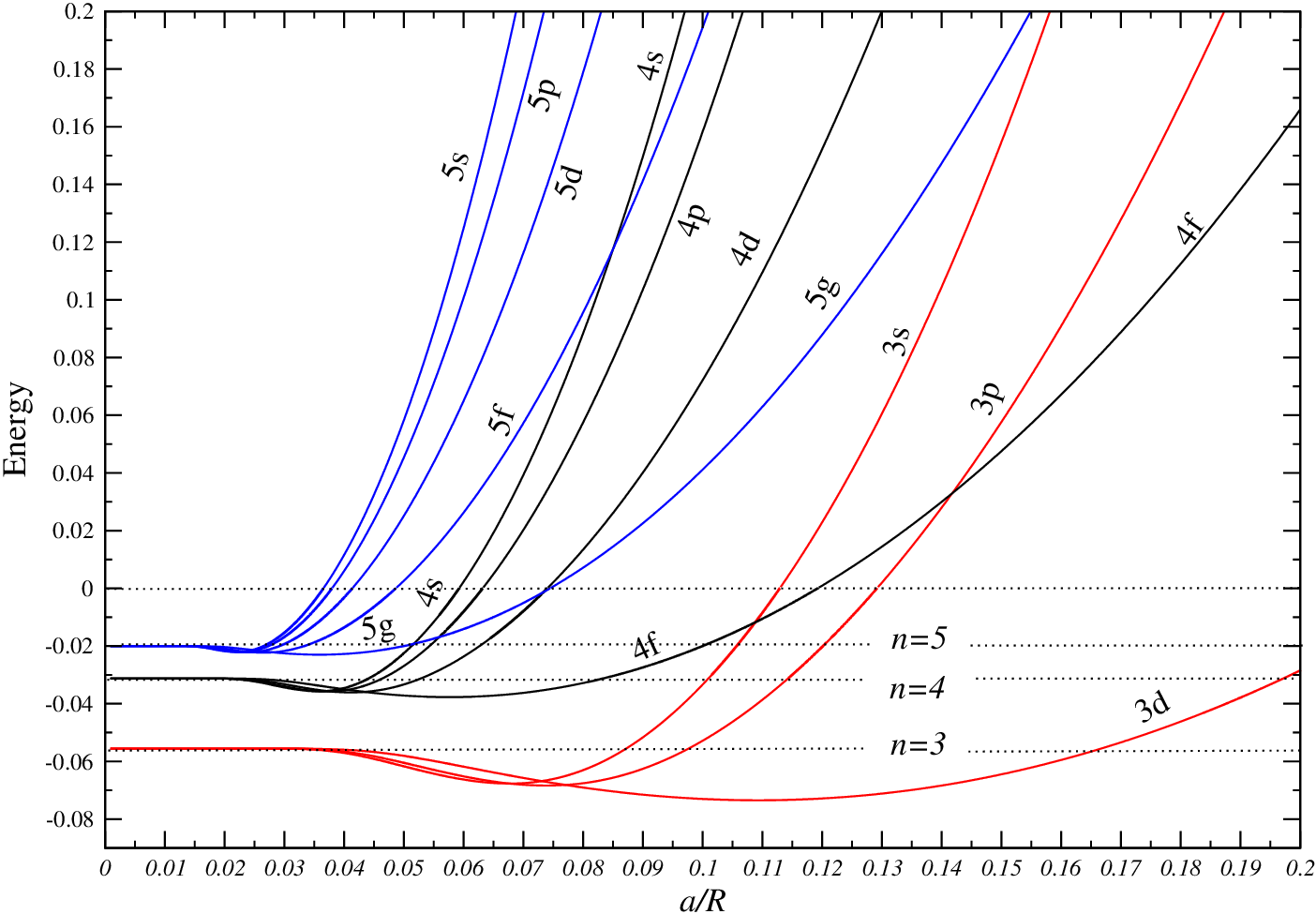,width=10cm}
\end{center}
\caption{\it Spectrum of a hydrogen atom centered in a spherical cavity with 
Neumann boundary condition (i.e.\ $\gamma = 0$) as a function of $a/R$.  The 
dotted lines represent the spectrum of the infinite system. Top: States with 
quantum numbers $n = 1, 2, 3$. Bottom: States with quantum numbers 
$n = 3, 4, 5$. The energy is given in units of $M e^4$.} 
\label{spec0}
\end{figure}
\begin{figure}[tbh]
\begin{center}
\epsfig{file=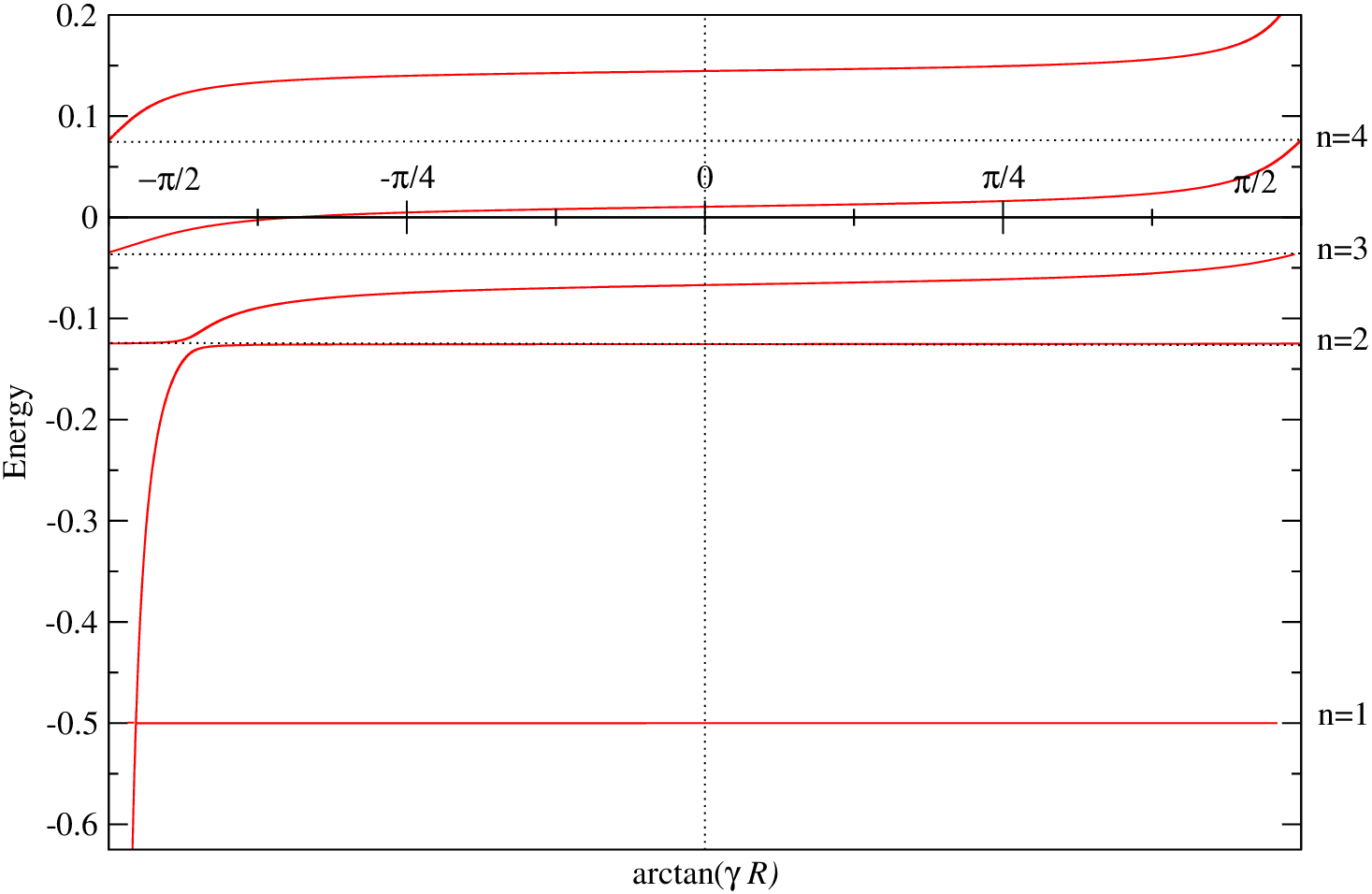,width=10cm} \vskip0.5cm
\epsfig{file=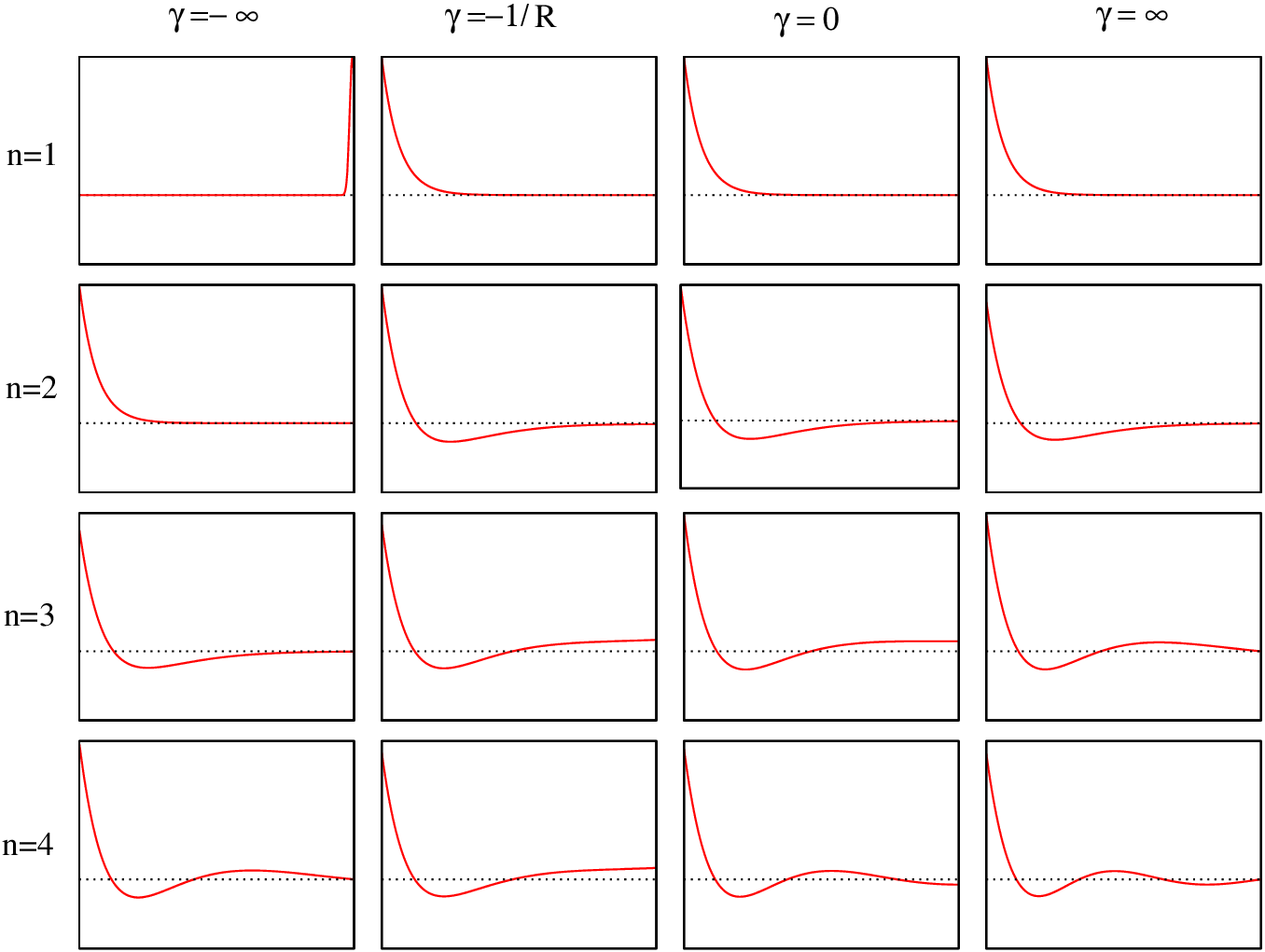,width=10cm}
\end{center}
\caption{\it Top: Spectrum of $l = 0$ states for a hydrogen atom in a spherical 
cavity of radius $R = 16 a$ with general Robin boundary conditions as a 
function of the self-adjoint extension parameter $\gamma$, rescaled to 
$arctan(\gamma R)$. The energy is given in units of $M e^4$.  The dotted 
lines represent the spectrum for $\gamma = \infty$. Bottom: Wave functions of 
the four lowest $l = 0$ states with $n = 1, 2, 3$ and 4 for 
$\gamma = \infty,0,-1/R$, and $-\infty$.}
\label{hydspec0}
\end{figure}
\begin{figure}[tbh]
\begin{center}
\epsfig{file=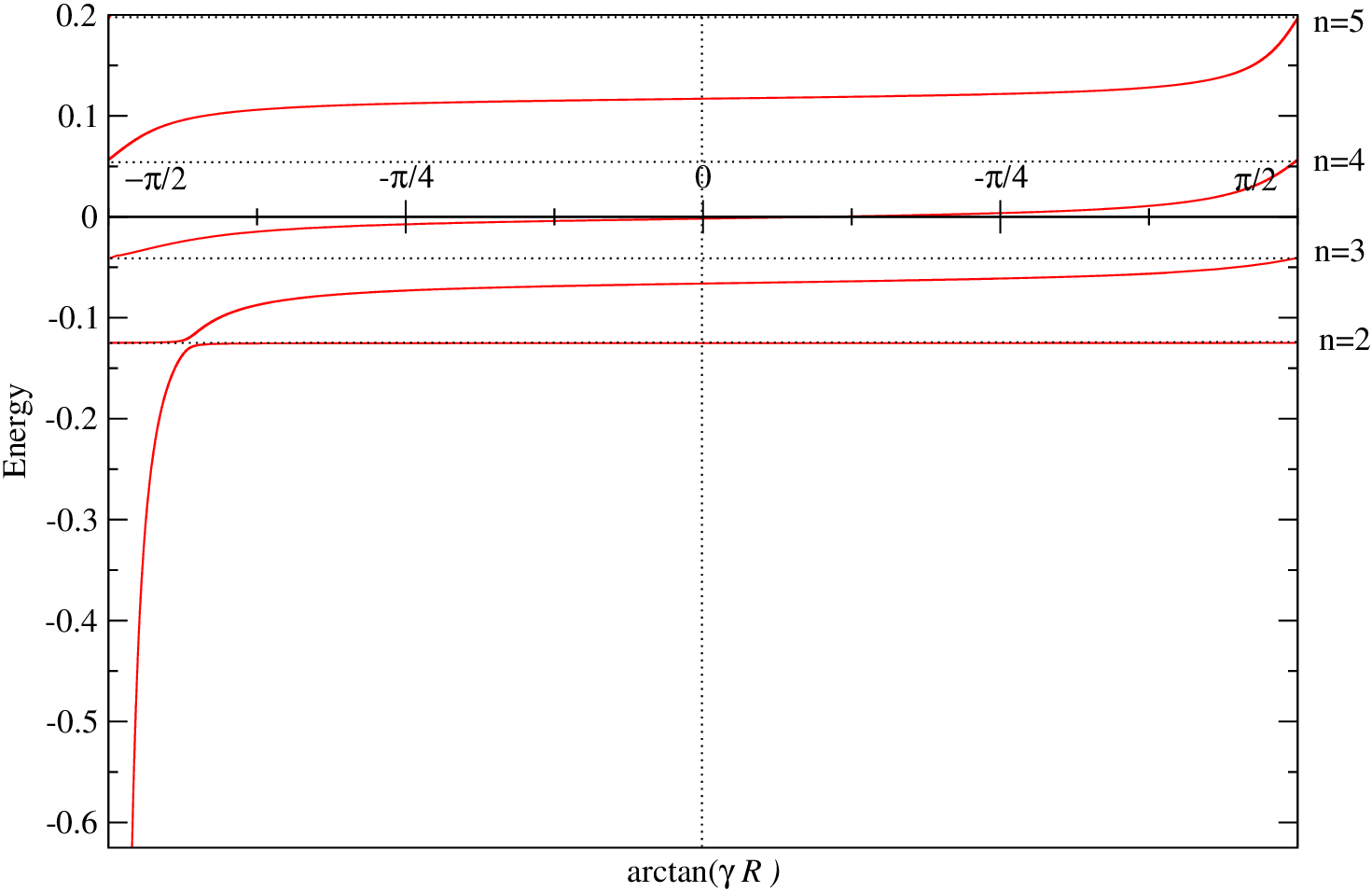,width=10cm} \vskip0.5cm
\epsfig{file=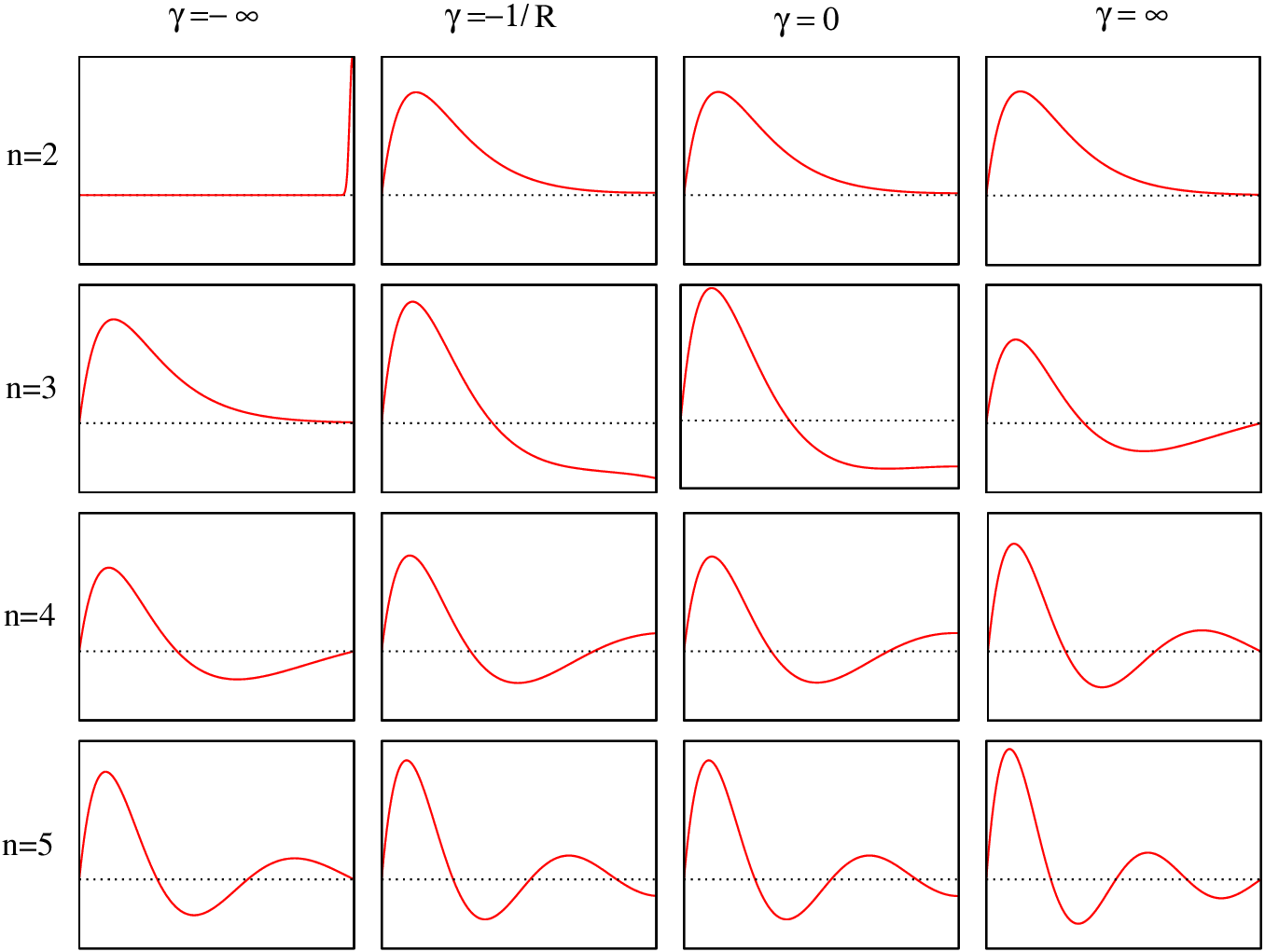,width=10cm}
\end{center}
\caption{\it Top: Spectrum of $l = 1$ states for a hydrogen atom in a spherical 
cavity of radius $R = 16 a$ with general Robin boundary conditions as a 
function of the self-adjoint extension parameter $\gamma$, rescaled to 
$arctan(\gamma R)$. The energy is given in units of $M e^4$. The dotted 
lines represent the spectrum for $\gamma = \infty$. Bottom: Wave functions of 
the four lowest $l = 1$ states with $n = 2, 3, 4$ and 5 for 
$\gamma = \infty,0,-1/R$, and $-\infty$.}
\label{hydspec1}
\end{figure}

\subsection{Energy spectrum}

\begin{figure}[tbh]
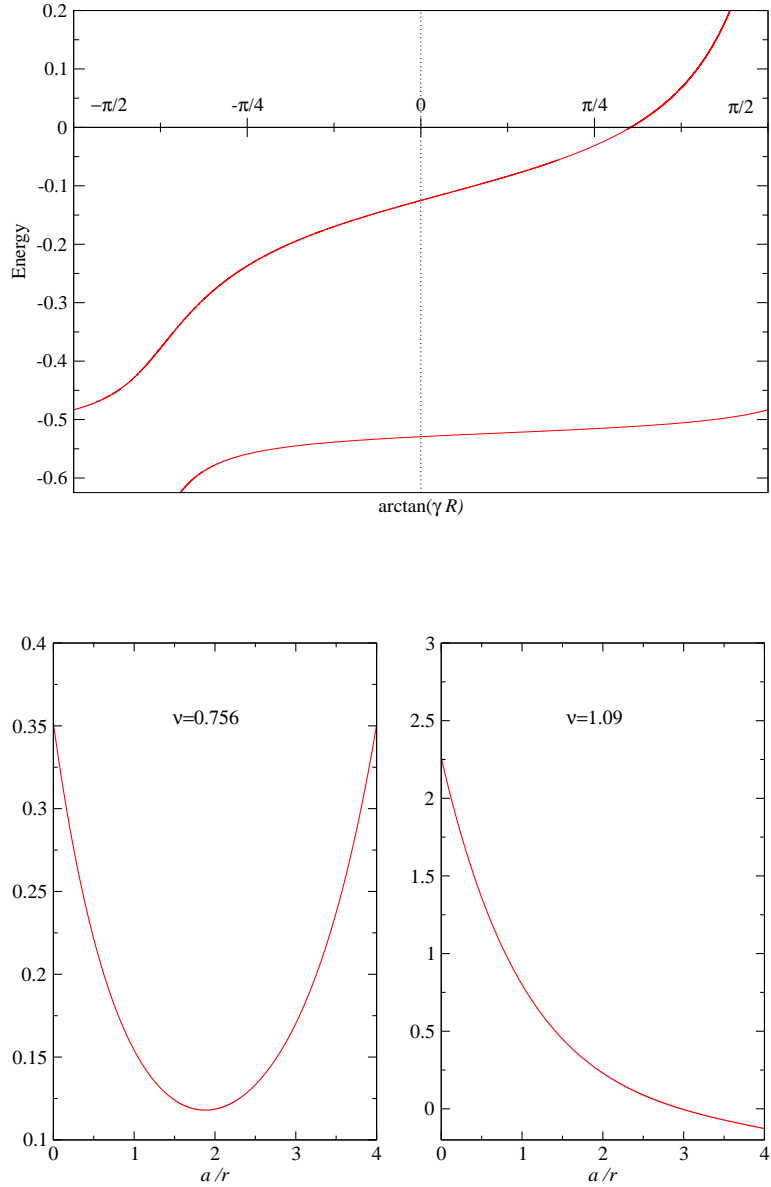

\begin{center}
\epsfig{file=hydfig9.eps,width=10cm} \vskip1.5cm
\epsfig{file=hydfig10.eps,width=10cm}
\end{center}
\caption{\it Top: Avoided level crossing between a 1s and a 2s state in a 
spherical cavity of radius $R = 4a$ indicating a cavity resonance. The energy 
is given in units of $M e^4$. Bottom: Wave functions of the two states for 
$\gamma = -0.8248/a$, which are localized both near the center and at the wall 
of the cavity.}
\label{resonance}
\end{figure}

We consider the Hamiltonian of the hydrogen atom,
\begin{equation}
H = - \frac{1}{2 M} \Delta - \frac{e^2}{r} = 
- \frac{1}{2 M} \left(\p_r^2 + \frac{2}{r} \p_r
- \frac{\vec L \, ^2}{r^2}\right) - \frac{e^2}{r}.
\end{equation}
Again, the wave function factorizes into
\begin{equation}
\Psi(\vec x) = \psi_{\nu l}(r) Y_{lm}(\theta,\varphi),
\end{equation}
and the radial equation now takes the form
\begin{equation}
\left[- \frac{1}{2 m} \left(\p_r^2 + \frac{2}{r} \p_r
- \frac{l(l + 1)}{r^2}\right) - \frac{e^2}{r}\right] \psi_{\nu l}(r) = 
E \psi_{\nu l}(r).
\end{equation}
In this case, we parameterize the energy as
\begin{equation}
E = - \frac{M e^4}{2 \nu^2}.
\end{equation}
While in the infinite volume the quantum number $\nu$ takes integer values for 
the bound state spectrum, in the cavity $\nu$ is in general real-valued. 
Introducing the Bohr radius
\begin{equation}
a = \frac{1}{M e^2},
\end{equation}
for negative energy the normalizable wave function is given by
\begin{equation}
\psi_{\nu l}(r) = A \left(\frac{2 r}{\nu a}\right)^l 
L^{2l+1}_{\nu-l-1}\left(\frac{2 r}{\nu a}\right) 
\exp\left(- \frac{r}{\nu a}\right),
\end{equation}
where $L^{2l+1}_{\nu-l-1}(2r/\nu a)$ is an associated Laguerre function. As 
before, for a spherical cavity with the most general perfectly reflecting 
boundary condition one obtains
\begin{equation}
\gamma \psi_{\nu l}(R) + \p_r  \psi_{\nu l}(R) = 0,
\end{equation}
The energy spectrum is then determined by the transcendental equation
\begin{equation}
\left(\frac{\gamma \nu a}{2} - \frac{1}{2} + \frac{l \nu a}{2 R}\right)
L^{2l+1}_{\nu-l-1}\left(\frac{2 R}{\nu a}\right) -
L^{2l+2}_{\nu-l-2}\left(\frac{2 R}{\nu a}\right) = 0.
\end{equation}
The finite volume effects on the energy spectrum are illustrated for the 
standard Dirichlet boundary condition (with $\gamma = \infty$) in figure
\ref{specinf}, and for Neumann boundary conditions (with $\gamma = 0$) in 
figure \ref{spec0}. In both cases, the accidental degeneracy between states of 
different angular momenta, which is generated by the Runge-Lenz vector, is 
lifted. While Dirichlet boundary conditions shift the energies upward, Neumann 
boundary conditions may lead to a downward shift of the energy. The energy 
spectrum for $l = 0$ as a function of the self-adjoint extension parameter 
$\gamma$ and the corresponding wave functions of the states with 
$n = 1, 2, 3, 4$ are illustrated in figure \ref{hydspec0}.
Analogous results are shown in figure \ref{hydspec1} for $l = 1$. In these
figures one notices some avoided level crossings, which correspond to hydrogen
bound states that resonate with states localized at the cavity wall. Such states
have been studied before by introducing additional potentials at the boundary
\cite{Con99,Con00}. Here cavity resonances emerge naturally from the Robin 
boundary condition. The fact that resonances in a finite volume manifest 
themselves as avoided level crossings is familiar from quantum field theory, in
particular, lattice field theory \cite{Wie89,Lue91}. Figure \ref{resonance} 
(top) zooms in on an avoided level crossing between a 1s and a 2s state in a 
spherical cavity of radius $R = 4a$. The corresponding wave functions are 
illustrated at the bottom of figure 
\ref{resonance}.

\subsection{Self-Adjointness of the Runge-Lenz Vector}

It is well known that the hydrogen atom in infinite space enjoys an accidental
$SO(4)$ symmetry generated by the angular momentum $\vec L$ together with the
Runge-Lenz vector
\begin{equation}
\vec R = \frac{1}{2 M} \left(\vec p \times \vec L - \vec L \times \vec p\right)
- \frac{e^2 \vec r}{r}.
\end{equation}
Due to this dynamically enhanced symmetry, the states with quantum number $n$
are $n^2$-fold degenerate. For example, the 2s state is degenerate with the
three 2p states, and the 3s state is degenerate with the three 3p and the five
3d states. As we have seen, the spectrum of hydrogen confined to a spherical
cavity no longer has these accidental degeneracies. Since the hard wall boundary
condition of the spherical cavity does not violate rotation invariance, the
angular momentum $\vec L$ is obviously still conserved. The Runge-Lenz vector, 
on the other hand, is no longer a conserved quantity. 

As differential operators, both the Hamiltonian and the Runge-Lenz vector still
have the same form as in the infinite volume. Hence, one might expect that they
still commute with each other. However, this is not the case, because the
action of the two operators is limited to the domain of the Hilbert space in
which they are self-adjoint. As we have shown, the Hamiltonian is self-adjoint
in the spherical cavity if the wave function obeys the boundary condition
eq.(\ref{bcdot}).  As we will now show, the Runge-Lenz vector is no longer 
self-adjoint in the domain of the Hamiltonian. This has already been noted by
Pupyshev and Scherbinin \cite{Pup98,Pup02} for the standard boundary condition
with $\gamma = \infty$. Here we extend their argument to arbitrary values of
$\gamma$. Thanks to rotation invariance, it is sufficient to restrict ourselves
to energy eigenstates $\Psi(\vec r) = \psi_{\nu l}(r) Y_{ll}(\theta,\varphi)$ 
with the maximum angular momentum projection $m = l$. As was shown in 
\cite{Pup98}, on these states the operator $R_+ = R_x + i R_y$ acts as
\begin{eqnarray}
\label{Runge}
R_+ \Psi(\vec r)&=&\left[\frac{l + 1}{M} \p_r \psi_{\nu l}(r) + 
\left(e^2 - \frac{l(l + 1)}{M r}\right) \psi_{\nu l}(r)\right] 
Y_{l+1,l+1}(\theta,\varphi) \nonumber \\
&=&\chi_{\nu,l+1}(r) Y_{l+1,l+1}(\theta,\varphi),
\end{eqnarray}
and thus $R_+$ raises the angular momentum from $l$ to $l+1$. In order to
investigate if $\vec R$ is self-adjoint, we must decide whether the new wave
function $\chi_{\nu,l+1}(r)$ obeys the boundary condition
\begin{equation}
\label{bcchi}
\gamma \chi_{\nu,l+1}(R) + \p_r \chi_{\nu,l+1}(R) = 0.
\end{equation}
By inserting the expression of eq.(\ref{Runge}) into this relation and by using 
the radial Schr\"odinger equation for $\psi_{\nu l}(r)$, one obtains
\begin{eqnarray}
\label{Rungeselfad}
&&\gamma \chi_{\nu,l+1}(R) + \p_r \chi_{\nu,l+1}(R) = \nonumber \\
&&\frac{l + 1}{M} \left[- \gamma \left(\gamma - \frac{2}{R}\right) +
\frac{l(l + 2)}{R^2} - \frac{2 M e^2}{R} - 2 M E \right] \psi_{\nu,l}(R).
\end{eqnarray}
Here we have used the boundary condition 
$\gamma \psi_{\nu,l}(R) + \p_r \psi_{\nu,l}(R) = 0$ for the original wave function.
Since $l + 1 \neq 0$, the right-hand side of eq.(\ref{Rungeselfad}) cannot 
vanish for all values of the energy $E$. Hence, the wave function 
$\chi_{\nu,l+1}(r)$ that results from the application of $R_+$ on $\psi_{\nu,l}(r)$
does not obey the boundary condition eq.(\ref{bcchi}), and thus does not lie in
the domain of the Hamiltonian. This implies that the Runge-Lenz vector $\vec R$
is not self-adjoint in this domain. Consequently, the fact that the differential
operators $H$ and $\vec R$ commute in the naive sense is irrelevant, because 
$H$ cannot even act on the wave function $\chi_{\nu,l+1}(r)$ generated by $R_+$,
since it is outside the domain of $H$. This explains why the accidental 
degeneracy associated with the Runge-Lenz vector is lifted in the spherical
cavity.
\begin{figure}[tbh]
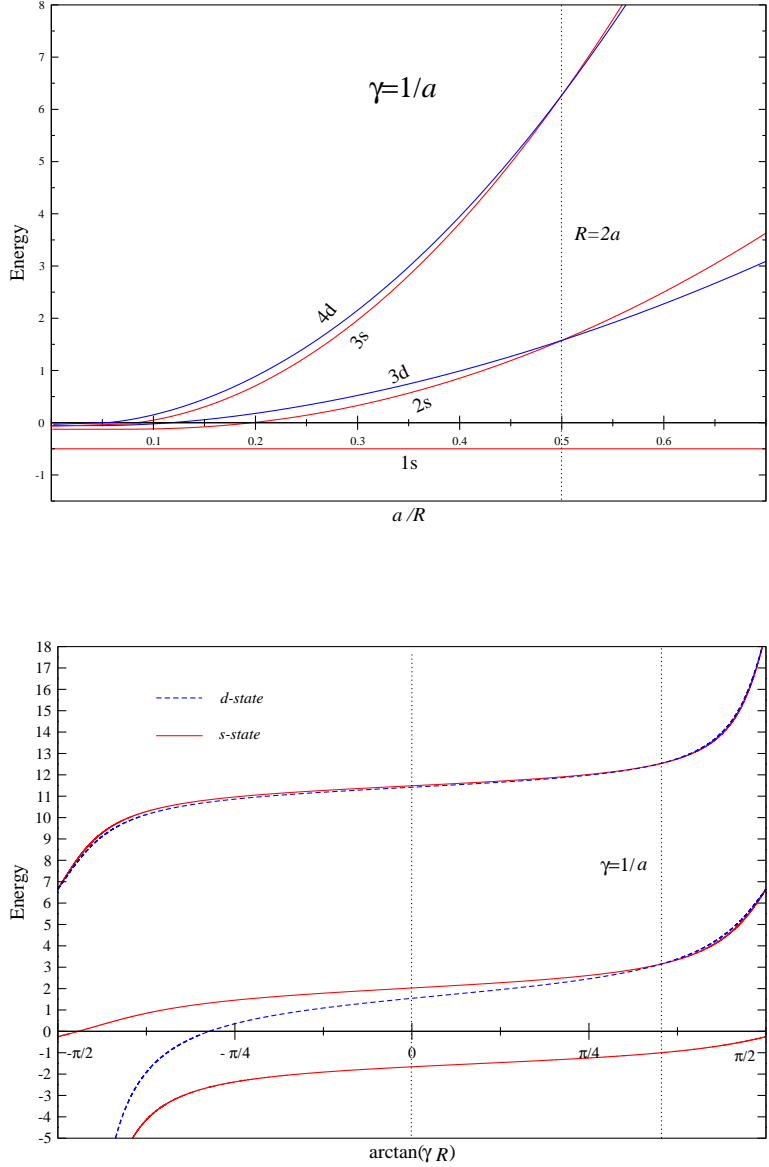

\begin{center}
\epsfig{file=hydfig11.eps,width=10cm} \vskip1.5cm
\epsfig{file=hydfig12.eps,width=10cm}
\end{center}
\caption{\it Top: Energy of s- and d-states for a hydrogen atom in a 
spherical cavity with $\gamma = 1/a$ as a function of $a/R$. There is an 
accidental degeneracy for $R = 2a$. Bottom: Energy of s- and d-states for a 
hydrogen atom in a spherical cavity with radius $R = 2a$ as a function of
$\arctan(\gamma R)$. The energies of the 3s and 4d states are very similar for 
all values of $\gamma$, but are identical only for $\gamma = 2/R = 1/a$ and for
$\gamma = \pm \infty$. The energy is given in units of $M e^4$. }
\label{accidental}
\end{figure}

Remarkably, as was pointed out in \cite{Pup98,Pup02}, at $\gamma = \infty$ a 
remnant of the accidental symmetry persists for a cavity of radius 
$R = (l + 1)(l + 2) a$. In that case, a state of angular momentum $l$ is 
degenerate with a state of angular momentum $l + 2$. For example, in a cavity
of radius $R = 2 a$, the states 2s and 3d, 3s and 4d, 4s and 5d, etc.\ are 
degenerate and thus form multiplets of $1 + 5 = 6$ degenerate states. Similarly,
for $R = 6 a$, the states 3p and 4f, 4p and 5f, 5p and 6f, etc.\ are 
degenerate, forming multiplets of $3 + 7 = 10$ degenerate states. These 
degeneracies arise because, for $R = (l + 1)(l + 2) a$, the operator $R_+^2$ 
maps a wave function $\psi_{\nu,l}(r)$ back into the domain of the Hamiltonian.
Again using the radial Schr\"odinger equation, it is straightforward to show
that $R_+^2 \psi_{\nu,l}(r) Y_{ll}(\theta,\varphi) = 
\chi_{\nu,l+2}(r) Y_{l+2,l+2}(\theta,\varphi)$ with
\begin{eqnarray}
\label{chieq}
\chi_{\nu,l+2}(r)&=& 
\frac{2 l + 3}{M} 
\left(e^2 - \frac{(l + 1)(l + 2)}{M r}\right) \p_r \psi_{\nu,l}(r) \nonumber \\
&+&\left[\frac{(l+1)(l+2)}{M}
\left(\frac{l(2l + 3)}{M r^2} - \frac{3 e^2}{r} - 2 E\right) +
e^2 \left(e^2 - \frac{l(l + 1)}{Mr}\right)\right] \psi_{\nu,l}(r). \nonumber \\
\,
\end{eqnarray}
For $\gamma = \infty$, the wave function obeys the boundary condition
$\psi_{\nu,l}(R) = 0$. Hence, for 
\begin{equation}
e^2 - \frac{(l + 1)(l + 2)}{M R} = 0 \ \Rightarrow \ 
R = \frac{(l + 1)(l + 2)}{M e^2} = (l + 1)(l + 2) a,
\end{equation}
the wave function $\chi_{\nu,l+2}(r)$ obeys the same boundary condition, 
$\chi_{\nu,l+2}(R) = 0$, and thus indeed lies in the domain of the Hamiltonian. 
Taking a derivative with respect to $r$ of eq.(\ref{chieq}), and once again 
using the radial
Schr\"odinger equation, one obtains an expression for $\p_r \chi_{\nu,l+2}(r)$. 
Combining this expression with eq.(\ref{chieq}) and using the general boundary 
condition $\gamma \psi_{\nu,l}(R) + \p_r \psi_{\nu,l}(R) = 0$, one finally obtains 
an expression for $\gamma \chi_{\nu,l+2}(R) + \p_r \chi_{\nu,l+2}(R)$, which is too
complicated to be displayed here. Remarkably, as was already pointed out in 
\cite{Sch00}, for $R = (l + 1)(l + 2) a$ and $\gamma = 2/R$ this expression 
again vanishes independent of the energy $E$, and thus $\chi_{\nu,l+2}(r)$ again
belongs to the domain of the Hamiltonian. Consequently, an additional accidental
degeneracy also arises in this case. For $R = 2 a$ and $\gamma = 2/R$ one finds 
that the states 2s and 3d, 3s and 4d, 4s and 5d, etc.\ are degenerate. This is
illustrated in figure \ref{accidental}. It is interesting to note that, in this
case, the energy of the 1s state is $R$-independent. This follows immediately
from the fact that the ground state wave function is a simple decaying 
exponential proportional to $\exp(- r/a)$. Similarly, for $R = 6 a$ 
and $\gamma = 2/R$, the states 3p and 4f, 4p and 5f, 5s and 6f, etc.\ are 
degenerate.

Following the same steps as before, one can check whether the radial wave 
function that results from $R_+^k \psi_{\nu,l}(r) Y_{ll}(\theta,\varphi) = 
\chi_{\nu,l+k}(r) Y_{l+k,l+k}(\theta,\varphi)$ obeys the self-adjoint extension 
condition $\gamma \chi_{\nu,l+k}(R) + \p_r \chi_{\nu,l+k}(R) = 0$ independent of 
the energy $E$. In a rather lengthy calculation we have convinced ourselves 
that this is not the case for $k = 3$ and 4, and we suspect that the same is 
true for larger values of $k$.

\section{Hydrogen Atom Confined to the Surface of a Finite Cone}

\begin{figure}[tbh]
\begin{center}
\epsfig{file=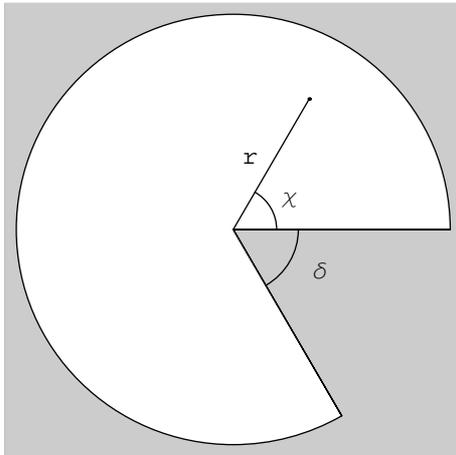,width=6cm}
\end{center}
\caption{\it A cone is obtained by cutting a wedge of deficit angle $\delta$ 
out of the 2-dimensional plane, and by gluing the open ends back together. 
Points on the cone are described by the distance $r$ from the tip and an angle 
$\chi$ which varies between $0$ and $2 \pi - \delta$.}
\label{cone}
\end{figure}

In \cite{AlH08} we have studied a particle confined to the surface of an
infinitely extended cone and bound to its tip by a $1/r$ potential. As 
illustrated in figure \ref{cone}, a cone is obtained from the plane by removing 
a wedge of deficit angle $\delta$ and gluing the open ends back together. As a 
consequence, the polar angle $\chi$ no longer extends from $0$ to $2 \pi$, but 
only to $2 \pi - \delta$. Similar to the hydrogen atom confined to a sphere, 
for deficit angles $\delta$ that are rational fractions of $2 \pi$, we found a 
remnant of the accidental $SU(2)$ symmetry generated by the Runge-Lenz vector 
$\vec R$. Remarkably, as a consequence of $\vec R$ not being self-adjoint in 
the domain of the Hamiltonian, some of the corresponding multiplets were found 
to have fractional ``spin'' and unusual degeneracies. 

In this section we extend the discussion to a particle confined to the surface
of a finite circular cone with general perfectly reflecting boundary conditions
characterized by the self-adjoint extension parameter $\gamma$. Again, the
particle is bound to the tip of the cone by a $1/r$ potential, i.e.\ we consider
a hydrogen atom with the proton residing at the tip of the cone. It is useful to
rescale the polar angle such that it again covers the full interval, i.e.\
\begin{equation}
\varphi = \frac{\chi}{s} \in [0,2 \pi],
\end{equation}
with the scale factor
\begin{equation}
s =  1 - \frac{\delta}{2 \pi}.
\end{equation}
The Hamiltonian then takes the form
\begin{equation}
H = - \frac{1}{2M} \left(\p_r^2 + \frac{1}{r} \p_r\right) -
\frac{1}{2 M r^2 s^2} \p_\varphi^2 - \frac{e^2}{r}.
\end{equation}
It should be noted that on the cone the centrifugal barrier is modified by the
additional factor $s^2$. 

Let us consider energy eigenstates $\Psi(\vec r) = \psi_{\nu m}(r) 
\exp(i m \varphi)$ with angular momentum quantum number $m \in \Z$.
In this case, the raising operator $R_+ = R_x + i R_y$ constructed from the
components of the Runge-Lenz vector acts as \cite{AlH08}
\begin{eqnarray}
R_+ \Psi(\vec r)&=&\frac{1}{M} \left[ 
- \left(\frac{m}{s} + \frac{1}{2}\right) \p_r + 
\frac{m}{s} \left(\frac{m}{s} + \frac{1}{2}\right) \frac{1}{r} - \frac{1}{a}
\right] \psi_{\nu m}(r) \exp(i (m + s) \varphi) \nonumber \\
&=&\chi_{\nu,m+s}(r) \exp(i (m + s) \varphi).
\end{eqnarray}
Unless $s \in \Z$, the resulting wave function is not $2 \pi$-periodic in
$\varphi$ and thus does not belong to the domain of the Hamiltonian. 
Consequently, the Runge-Lenz vector is in general not self-adjoint in that
domain, and hence it does not generate a proper accidental symmetry. Still, if
the deficit angle $\delta$ is a rational fraction of $2 \pi$ (i.e.\ if $s = p/q$
with $p, q \in \Z$), $q$ applications of $R_+$ lead back into the domain of
the Hamiltonian, and hence a remnant of the accidental symmetry is still 
present. Here we restrict ourselves to two applications of $R_+$, which yields
\begin{eqnarray}
R_+^2 \Psi(\vec r)&=&\frac{1}{M^2} \left[ 
- \left(\frac{m}{s} + \frac{3}{2}\right) \p_r + 
\left(\frac{m}{s} + 1\right) \left(\frac{m}{s} + \frac{3}{2}\right) \frac{1}{r} 
- \frac{1}{a} \right] \nonumber \\
&\times&\left[- \left(\frac{m}{s} + \frac{1}{2}\right) \p_r + 
\frac{m}{s} \left(\frac{m}{s} + \frac{1}{2}\right) \frac{1}{r} - \frac{1}{a}
\right] \psi_{\nu m}(r) \exp(i (m + 2 s) \varphi) \nonumber \\
&=&\chi_{\nu,m+2s}(r) \exp(i (m + 2s) \varphi).
\end{eqnarray}
In this case, if $s$ is an integer or half-integer, the resulting wave function 
is indeed periodic. However, this alone does not guarantee that it belongs to 
the domain of the Hamiltonian. For this to be the case, the wave function must 
also obey the boundary condition 
$\gamma \chi_{\nu,m+2s}(R) + \p_r \chi_{\nu,m+2s}(R) = 0$. In complete analogy
to the case of the sphere, it is straightforward to convince oneself that this
relation is indeed satisfied for 
\begin{equation}
R = \left[\left(\frac{m}{s} + 1\right)^2 - \frac{1}{4}\right] a,
\end{equation}
provided that the self-adjoint extension parameter takes one of the three values
$\gamma = 3/(2R)$ or $\pm \infty$. Hence, if $2s \in \Z$, even on a finite cone 
a remnant of the accidental symmetry generated by the Runge-Lenz vector 
persists. In this case, the accidental symmetry gives rise to the degeneracy of 
states with angular momentum $m$ and $m + 2s$. 

\section{Conclusions}

We have investigated both a free particle and an electron bound in a hydrogen 
atom confined to a spherical cavity with general perfectly reflecting boundary 
conditions characterized by a self-adjoint extension parameter $\gamma$.
For negative values of $\gamma$, bound states localized at the cavity wall may
arise. In particular, a ``free'' particle (which seems to have only kinetic 
energy) may then
have negative energy, since it can bind to the wall. While such states seem to 
violate the Heisenberg uncertainty relation, we have explicitly verified that
they are perfectly consistent with a properly generalized uncertainty relation
that we have recently derived for a finite volume. 

When a hydrogen atom is placed at the center of a spherical cavity, the 
accidental degeneracy of states with different angular momenta $l$ is lifted. 
At the classical level, this manifests itself by the fact that the particle's 
orbit is in general no longer closed. At the quantum level, the Runge-Lenz 
vector $\vec R$ ceases to be self-adjoint in the domain of the Hamiltonian. 
Remarkably, for the specific value $R = (l+1)(l+2) a$ of the cavity radius and 
for $\gamma = 2/R$, or $\pm \infty$, the operator $R_+^2 = (R_x + i R_y)^2$, 
which
turns states with angular momentum $l$ into states with angular momentum $l+2$,
still commutes with the Hamiltonian. This gives rise to accidental degeneracies
(e.g.\ of s and d-states) even in a finite volume. The same is true for an
electron confined to the surface of a finite circular cone of radius $R$ with
deficit angle $\delta = \pi$, and bound to its tip by a $1/r$ potential. In 
that case, there are accidental degeneracies for $R = [(2 m + 1)^2 - 1/4]a$ and
$\gamma = 3/(2R)$ or $\pm \infty$ between states of angular momentum $L_z = m$ 
and $m + 1$. Investigating these degeneracies in more detail seems worthwhile 
for future studies. 

For specific values of the confinement radius $R$ or of the self-adjoint 
extension parameter $\gamma < 0$, a hydrogen bound state may resonate with a
bound state localized at the cavity wall. This leads to patterns of avoided
level crossings in the energy spectrum. Similar patterns arise when bound states
localized on the wall are modeled with additional attractive potentials
\cite{Con99,Con00}. The modeling with a non-trivial self-adjoint extension 
parameter is mathematically more appealing. It is plausible that, e.g., the 
description of atoms encapsulated in fullerenes can benefit from this method. 
Not only for this reason, it remains promising to further investigate the 
self-adjoint extensions of quantum mechanical Hamiltonians.

\section*{Acknowledgments}

This work is supported in parts by the Schweizerischer Nationalfonds (SNF).
M.\ H.\ A.\ also likes to thank the city of Bern for supporting him in the 
framework of the Swiss national qualification program 
Bio\-medi\-zin-Na\-tur\-wis\-sen\-schaft-For\-schung (BNF). The ``Albert 
Einstein Center for Fundamental Physics'' at Bern University is supported by 
the ``Innovations- und Kooperationsprojekt C-13'' of the Schweizerische 
Uni\-ver\-si\-t\"ats\-kon\-fe\-renz (SUK/CRUS).

\end{document}